\documentclass[%
 reprint,
 amsmath,amssymb,
 aps, pra,
floatfix,
]{revtex4-2}

\usepackage{graphicx}
\usepackage{dcolumn}
\usepackage{bm}
\usepackage{hyperref}
\usepackage[mathlines]{lineno}
\usepackage{mathtools}
\usepackage{multirow}
\usepackage{adjustbox}
\usepackage{xcolor} 
\usepackage{makecell}

\newcommand{\D}{\Delta^{2}}
\newcommand{\hd}{\hat{\delta}}
\newcommand{\hx}{\hat{x}}
\newcommand{\hp}{\hat{p}}
\newcommand{\hq}{\hat{q}}
\newcommand{\Rom}[1]{\uppercase\expandafter{\romannumeral #1\relax}}

\begin{document}
\preprint{APS/123-QED}

\title{Camera-enabled scalable homodyne detection of multimode quantum light}
\author{Young-Do Yoon, Chan Roh, Geunhee Gwak}
\author{Young-Sik Ra}
\email{youngsikra@gmail.com}
\affiliation{Department of Physics, Korea Advanced Institute of Science and Technology (KAIST), Daejeon, 34141, Korea}
\date{\today}

\begin{abstract}
Scalability is a key challenge in advancing quantum technologies such as quantum computing, communication, and metrology. Photonic systems offer a promising route to scalability by enabling the deterministic generation of large-scale entangled states. Homodyne detection is an essential quantum measurement to exploit such entangled states, enabling quantum-enhanced measurement, deterministic quantum teleportation, GKP-state breeding, and quantum error correction. Despite the recent progress in generating large-scale quantum states, realizing quantum measurement at scale remains a major challenge. Here we realize scalable and efficient homodyne detection by leveraging a large number of pixels in a charge-coupled-device (CCD) camera. Our approach enables shot-noise-limited quadrature measurements of 60 optical modes simultaneously, while requiring only nanowatt-level local oscillator power per mode---a six-order-of-magnitude reduction compared to conventional methods. The system achieves clearance exceeding 24 dB for all modes with negligible crosstalk. We demonstrate its compatibility with a large-scale quantum state by directly observing squeezing and entanglement in 60 optical modes. Furthermore, we showcase applications in verifying multipartite entanglement and in the conditional preparation of multimode states. This work provides a scalable method for quantum measurement, paving the way for large-scale quantum information processing.
\end{abstract}
\maketitle

Scalability is widely recognized as a key requirement for realizing the full potential of quantum information technologies. Continuous-variable quantum optical systems offer a promising route to scalability by enabling the deterministic generation of large-scale quantum entanglement \cite{Asavanant2019, Larsen2019, Aghaee_Rad2025, Roh2025, Zhong:2020aa, Madsen2022}. Scalable generation of squeezed light and its interaction via beam splitter networks have enabled both cluster-state generation for measurement-based quantum computing \cite{Asavanant2019, Larsen2019, Aghaee_Rad2025, Roh2025} and Gaussian Boson Sampling for quantum simulation \cite{Zhong:2020aa, Madsen2022}. More recently, these techniques have advanced towards the generation of quantum states for quantum error correction, including Gottesman–Kitaev–Preskill (GKP) states \cite{Konno:2024aa, Larsen:2025aa} and high-dimensional cluster states \cite{Roh2025}. In contrast to these advances on the state-generation side---across both bulk \cite{Asavanant2019, Larsen2019, Zhong:2020aa, Madsen2022, Chen:2014jx, Barakat2025, Roh2025} and integrated \cite{Nehra:2022in, Aghaee_Rad2025, Larsen:2025aa, Jia:2025aa} optics---the scalable implementation of quantum measurement remains challenging. Moreover, most efforts have focused on photon-number detection \cite{Eaton:2023aa, Oripov:2023aa, Aghaee_Rad2025, Larsen:2025aa}, although homodyne detection plays a central role in continuous-variable quantum information processing \cite{Weedbrook:2012fe, Fabre:2020aa}.

Homodyne detection measures the field quadratures of quantum light, providing access to its wave-like properties \cite{Weedbrook:2012fe, Fabre:2020aa}. Homodyne detection is essential for harnessing the nonclassical noise reduction offered by squeezed light, enabling quantum-enhanced measurement \cite{Guo:2019el, Casacio2021, Jia:2024aa}. It also facilitates deterministic Bell-state measurements for quantum teleportation \cite{Takeda:2013hn, Takeda:2019aa}. Additional applications include entanglement distillation \cite{Dong:2008cj}, quantum key distribution \cite{Grosshans:2003aa}, and quantum random number generation \cite{Gabriel:2010gb}. Essential for quantum metrology and communication, homodyne detection becomes indispensable for quantum computing. As a key element of universal quantum computing \cite{Menicucci:2006ir}, it is required for shaping cluster states for measurement-based quantum computing \cite{Miwa:2010ek, Larsen2021} and for deterministically implementing non-Gaussian gates via quantum teleportation \cite{Takeda:2019aa}. Moreover, it is crucial for generating and breeding GKP states \cite{Konno:2024aa, Aghaee_Rad2025, Larsen:2025aa}, as well as for performing syndrome measurements for quantum error correction \cite{Gottesman:2001jb}.

Despite its importance, the scalable implementation of homodyne detection faces significant challenges. It inherently requires a local oscillator (LO) as a phase reference and a signal amplifier, and the LO power must be sufficiently high to ensure a high signal-to-noise ratio, known as clearance \cite{Appel:2007hn}. While a single homodyne detector requires only moderate laser power (tens of milliwatts for 20 dB clearance \cite{Aoki2009, Armstrong2009, Cai2021, Tasker:2020cv, Jia:2025aa}), scaling to fifty detectors demands nearly a watt of power. Moreover, severe circuit crosstalk and the stringent requirement for low electronic noise further complicate the implementation \cite{Armstrong2009, Cai2021}, and reliable detection of quantum properties additionally requires high quantum efficiency~\cite{Asavanant2019, Larsen2019, Aghaee_Rad2025, Roh2025, Konno:2024aa, Larsen:2025aa}. For these reasons, achieving scalable and efficient homodyne detection remains challenging \cite{Aoki2009, Armstrong2009, Cai2021}, even on photonic chips \cite{Tasker:2020cv, Aghaee_Rad2025, Jia:2025aa}.

We address these challenges by introducing a CCD camera for homodyne detection. In quantum optics, cameras have typically been employed for detecting single photons \cite{Lemos:2014aa, Chrapkiewicz2016, Zia:2023aa}, measuring intensity correlations \cite{Brida2010, Edgar2012, Gregory2020}, or estimating the variance of optical noise \cite{Cuozzo2022, Barakat2025, Kalash2025}. In this work, we develop a camera-based homodyne detection technique that operates in the shot-noise-limited regime---i.e., with classical noise fully suppressed---and measures individual quadrature outcomes for multiple modes, as required for quantum information processing. A key advantage is the ability to exploit a large number of camera pixels with high quantum efficiency, enabling scalable and efficient homodyne detection. Our method requires a significantly reduced LO power---only a few nanowatts---representing a six-orders-of-magnitude reduction over conventional approaches \cite{Aoki2009, Armstrong2009, Cai2021, Tasker:2020cv, Jia:2025aa}. This low-power operation offers two main advantages: (i) scalability, where even a few milliwatts of laser power, for example, suffice to operate over a million homodyne detectors; and (ii) noise robustness, as low-power lasers naturally exhibit reduced classical noise, favorable for shot-noise-limited operation. With these advantages, we demonstrate simultaneous homodyne detection of 60 optical modes using a one-dimensional camera array. We achieve a clearance exceeding 24~dB using less than 2~nW laser power per mode, with negligible crosstalk between different modes.

Our entire quantum system integrates this scalable detection unit with a large-scale quantum light source \cite{Roh2025, Gwak2024}, enabling full-system operation. We demonstrate the compatibility of the camera-based multimode homodyne detection by directly observing nonclassical noise reduction and entanglement in 60 optical modes. Moreover, we exploit the unique reconfigurability of multimode homodyne detection, which enables quadrature measurements in a superposed mode basis without requiring additional interferometers \cite{Weedbrook:2012fe, Fabre:2020aa}. With this capability, we demonstrate direct observation of multipartite entanglement without altering the experimental setup. Finally, we show an application of multimode homodyne detection to the conditional preparation of multimode states.

\section{Multimode homodyne detection}
\begin{figure*}[!ht]
	\includegraphics[width=\linewidth]{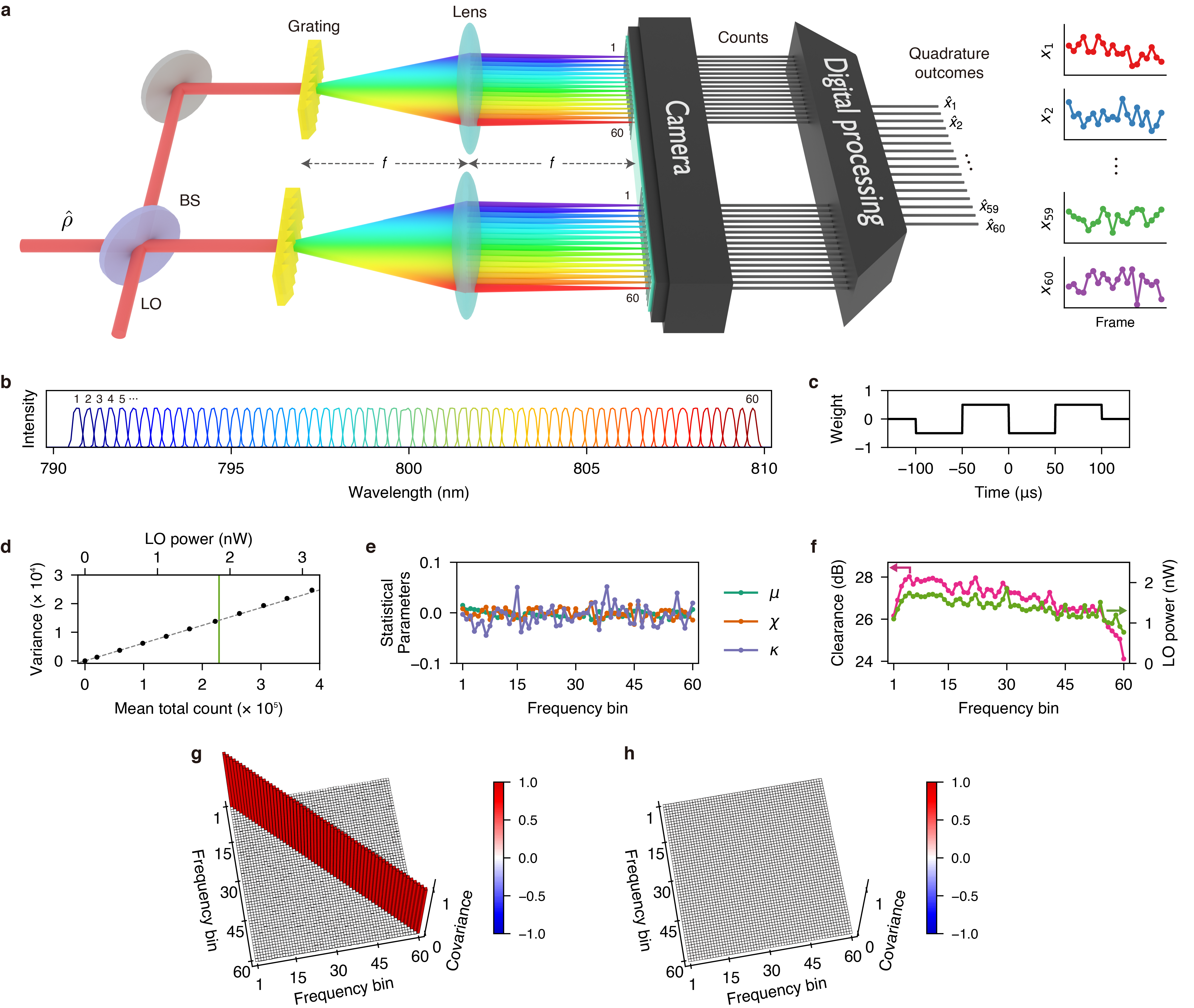}
	\centering
	\caption{\label{fig:setup} \textbf{Camera-based homodyne detection.}
    \textbf{a}, Experimental scheme. A quantum state of light in multiple frequency modes, $\hat{\rho}$, is interfered with a local oscillator (LO) at a 50:50 beam splitter (BS). The two output beams are diffracted by gratings and focused by lenses in a $2f$ configuration. As a result, a corresponding pair of camera pixels---one from each output beam and located on the left and right halves of the camera---maps each frequency-bin mode in \textbf{b}. During digital processing, the camera counts from each pixel pair---after gain balancing and weighted summation using the weight function in \textbf{c}---are subtracted and then normalized. Extending this procedure to all pixel pairs yields the quadrature outcomes of homodyne detection for 60 frequency-bin modes in each frame. The rightmost plots in \textbf{a} show example traces of the 60 quadrature outcomes.
    \textbf{b}, 60 frequency-bin modes and \textbf{c}, the weight function used in the experiment.
    \textbf{d}, Linearity test for homodyne detection at the 30th frequency-bin mode. By varying the mean total count of the associated pixel pair in each frame, we plot the variance of their gain-balanced count difference, weighted by the weight function defined in \textbf{c}. The data are shown after subtraction of the electronic-noise contribution. The dashed line shows the linear component of a quadratic fit to the data, and the green vertical line indicates the LO power used for homodyne detection.
    \textbf{e}, Offset ($\mu$), skewness ($\xi$), and excess kurtosis ($\kappa$) of the shot noise distribution. All these parameters are close to zero, following the expected Gaussian statistics.
    \textbf{f}, Clearance and LO power for each frequency-bin mode. All clearances exceed 24 dB, requiring only a few nanowatts LO power per mode. 
    \textbf{g}, Covariance matrix of the quadrature outcomes for the shot noise (i.e., the measurement of the vacuum state), and \textbf{h}, that of the electronic noise. The covariance matrix of the shot noise is nearly the identity matrix, and that of the electronic noise is close to the zero matrix.}
\end{figure*}

Our camera-based homodyne detection is illustrated in Fig.~\ref{fig:setup}a. For the LO, we use a femtosecond laser (75 fs duration, 800 nm central wavelength) that delivers a broadband spectrum for multimode homodyne detection. Its spectral amplitude and phase are flattened using a pulse shaper, supporting a wavelength range from 790 nm to 810 nm. At a high-efficiency CCD camera (Teledyne BLAZE 400HRX with 95\% quantum efficiency), 60 distinct pixel pairs register different frequency-bin modes, as shown in Fig.~\ref{fig:setup}b, with an average spectral overlap between neighboring bins of just 1.4\%. During digital processing, the counts from each pixel pair are first weighted by the weight function in Fig.~\ref{fig:setup}c, then gain-balanced and subtracted, and finally normalized such that the vacuum variance equals one. This procedure yields simultaneous quadrature outcomes for 60 modes, as illustrated in the rightmost plots in Fig.~\ref{fig:setup}a. More details of the experiment are provided in Supplementary Section \Rom{1}--\Rom{3}.

To benchmark the performance of the camera-based homodyne detector, we use an input state $\hat{\rho}$ prepared in the vacuum state. We first examine the detector's linearity, by comparing the mean total count of each pixel pair in a single frame with the variance of their gain-balanced count difference, weighted by the weight function in Fig.~\ref{fig:setup}c. As an example, the result for the 30th frequency-bin mode is shown in Fig.~\ref{fig:setup}d. The experimental data are fitted with a quadratic function, and its linear component is shown as the dashed line. The data follow the linear trend, as expected for shot-noise-limited detection. The coefficient of determination ($R^2$) between the linear component and the data is $0.996$, and the contribution of the linear component ($\eta_L$) from the quadratic fit at the LO power used is $0.985$. All 60 modes exhibit similar linear behavior as shown in Fig.~\ref{fig:linear_raw}. We further examine the Gaussianity of the quadrature outcomes: the offset, skewness, and excess kurtosis are all found to be nearly zero, as shown in Fig.~\ref{fig:setup}e.

Having confirmed shot-noise-limited operation, we next evaluate the clearance of the homodyne detection. The clearance is defined as the ratio of the shot-noise variance to the electronic-noise variance~\cite{Appel:2007hn}. Figure~\ref{fig:setup}f shows high clearance across all 60 modes, ranging from 24~dB to 28~dB (corresponding to 99.6--99.8\% efficiency~\cite{Appel:2007hn}). Notably, the required LO power is remarkably low---less than 2~nW---thanks to the low readout noise and dark noise of the CCD camera. This extremely low LO power is advantageous for achieving shot-noise-limited detection and scaling homodyne measurements to a large number of modes. We also investigate mode crosstalk in the homodyne detector. The covariance matrix of the shot noise (Fig.~\ref{fig:setup}g) shows negligible crosstalk, and the electronic noise (Fig.~\ref{fig:setup}h) is substantially smaller, confirming the reliable operation of our detector for multimode homodyne detection. Further details of the benchmarking method are provided in Supplementary Section \Rom{4}.

\section{Detecting multimode quantum light}

\begin{figure*}[htbp]
	\includegraphics[width=\linewidth]{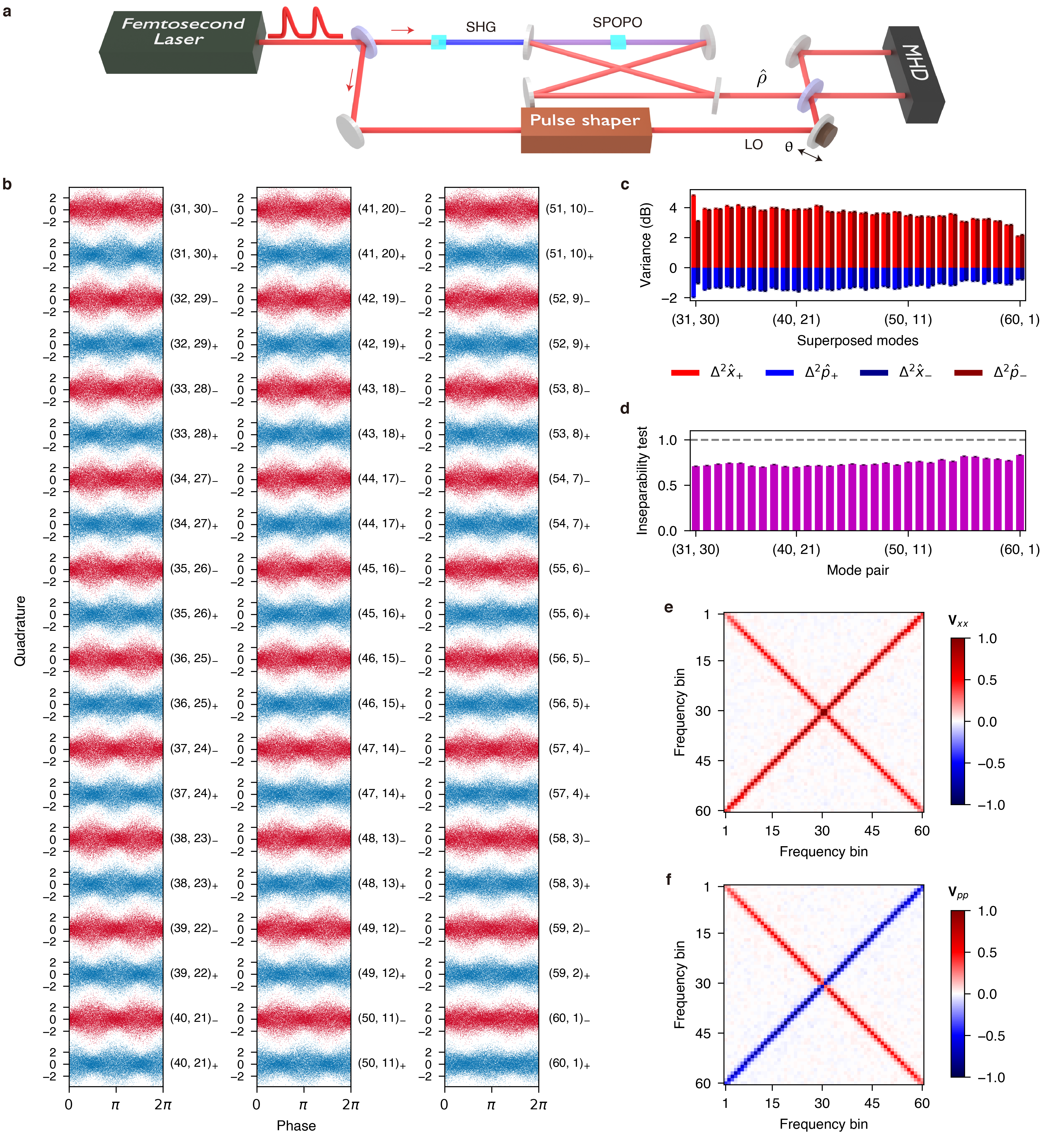}
	\centering
	\caption{\label{fig:epr}\textbf{Homodyne detection of multimode quantum light.}
    \textbf{a}, Experimental scheme. A multimode quantum state $\hat{\rho}$, generated from a synchronously pumped optical parametric oscillator (SPOPO), is measured using a camera-based multimode homodyne detector (MHD).
    Quadrature outcomes for the 60 frequency-bin modes in Fig.~\ref{fig:setup}b are simultaneously obtained as varying the LO phase $\theta$.
    SHG: second harmonic generation.
    In \textbf{b-d}, we consider 30 pairs of the frequency-bin modes, $(30+k,31-k)$ for $1\le k \le 30$.
    \textbf{b}, Quadrature outcomes for symmetric (blue, $(30+k,31-k)_+$) and antisymmetric (red, $(30+k,31-k)_-$) superpositions of two frequency-bin modes. These outcomes are obtained by adding (the symmetric case) or subtracting (the antisymmetric case) the quadrature outcomes of the frequency-bin modes. Each plot shows 24,000 quadrature outcomes. 
    \textbf{c}, Variances of $\hx$- and $\hp$-quadratures for symmetric ($\Delta^2\hx_{+}, \Delta^2\hp_{+}$) and antisymmetric ($\Delta^2\hx_{-}, \Delta^2\hp_{-}$) superpositions in \textbf{b}.
    \textbf{d}, Duan inseparability test for the 30 pairs, where the dashed line represents the classical bound. All pairs exhibit inseparability.
    \textbf{e} and \textbf{f}, Covariance matrices of $\hx$ quadratures (\textbf{e}) and $\hp$ quadratures (\textbf{f}) for the 60 frequency-bin modes, where the identity matrix has been subtracted for clarity.
    Error bars in \textbf{c} and \textbf{d} represent $\pm 1\sigma$ of the experimental data, calculated as the standard deviation of the sample variance assuming a Gaussian distribution.
    }
\end{figure*}

Next, we explore the capability of the camera-based homodyne detection for measuring highly multimode quantum light. Figure~\ref{fig:epr}a illustrates the experimental setup. A synchronously pumped optical parametric oscillator (SPOPO) generates a quantum state of light over multiple frequency modes~\cite{Roh2025, Roslund2014, Gwak2024}. Using the camera-based homodyne detector, we simultaneously measure the 60 frequency-bin modes of the generated state as varying the phase of the LO. 
See Supplementary Information for more details of the experimental setup (Section \Rom{1}) and for the LO phase estimation (Section \Rom{5}). Quadrature outcomes from the 60 frequency-bin modes exhibit random thermal noise, as shown in Fig.~\ref{fig:quad_raw}. However, these quadrature outcomes are actually highly correlated. In Fig.~\ref{fig:epr}b, we observe phase-dependent noise when the quadrature outcomes are added (blue) or subtracted (red) for the mode pair $(30+k,31-k)$, where $1\le k \le 30$. This is because the addition and subtraction of the quadrature outcomes correspond to the quadrature outcomes of the symmetrically and antisymmetrically superposed modes, respectively, which contain squeezed vacua~\cite{Roh2025, Roslund2014}. The variances of the $\hx$- and $\hp$-quadrature outcomes in these superposed modes, each estimated from 16,000 outcomes, are shown in Fig.~\ref{fig:epr}c. All of the 60 superposed modes exhibit nonclassical noise reduction below the vacuum noise, characterized by $\Delta^2\hp_{+}$ for symmetric and $\Delta^2\hx_{-}$ for antisymmetric superpositions. A small deviation of the overall trend in $(31,30)$ is due to the nonzero overlap between the adjacent modes.

We further verify the entanglement of each mode pair using the Duan inseparability criterion~\cite{Duan2000}, $\left( \D \hx_{-} + \D \hp_{+} \right) / 2 < 1$. As shown in Fig.~\ref{fig:epr}d, all pairs pass the entanglement test. Lastly, we directly investigate the correlations among the quadrature outcomes of the 60 frequency-bin modes (Fig.~\ref{fig:epr}e and f), using 16,000 outcomes for each of the $\hx$- and $\hp$-quadrature measurements. As expected, $\hx$-quadrature covariance matrix exhibits positive correlations, whereas $\hp$-quadrature covariance matrix shows negative correlations~\cite{Roh2025, Roslund2014}.

\section{Detecting multipartite entanglement}

\begin{figure*}[htbp]
	\includegraphics[width=\linewidth]{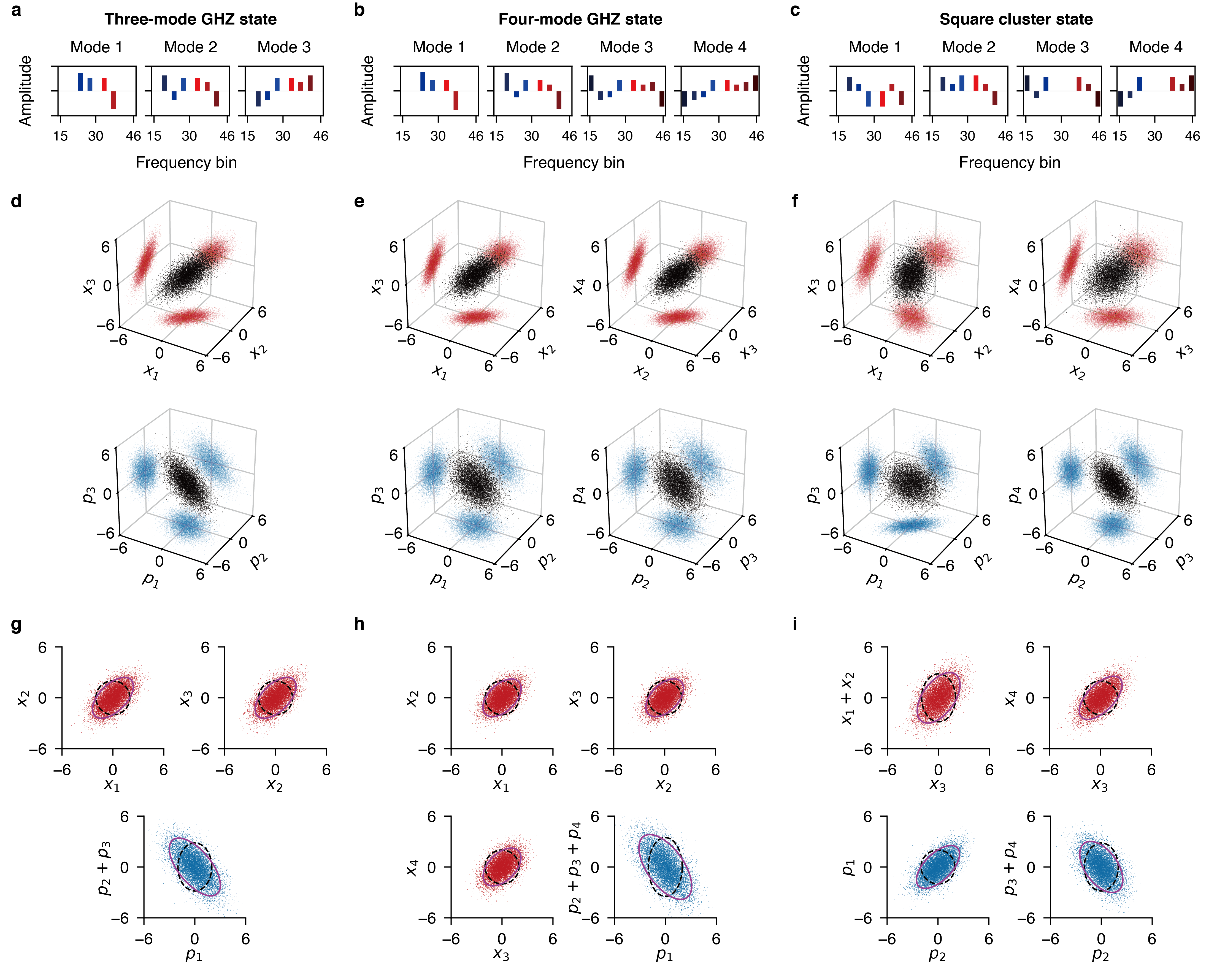}
	\centering
	\caption{\label{fig:multimode_state} \textbf{Homodyne detection of multipartite entangled states.} Left (\textbf{a},\textbf{d},\textbf{g}): a three-mode GHZ state, Middle (\textbf{b},\textbf{e},\textbf{h}): a four-mode GHZ state, Right (\textbf{c},\textbf{f},\textbf{i}): a (locally equivalent) four-mode square cluster state. \textbf{a-c}, Frequency-mode basis for each state. 
    We use the frequency bins in Fig.~\ref{fig:setup}\textbf{b}, grouping each pair of adjacent bins from the 15th to the 46th into a single mode, while discarding alternate pairs to improve spectral mode separation.
    \textbf{d-f}, Correlations of three different quadrature measurements (black dots). Two-dimensional projections are shown as red dots for the $\hx$ quadrature and blue dots for the $\hp$ quadrature.
    \textbf{g-i}, Quadrature correlations used to identify nullifier noise. The nullifiers for each state are defined in Table~\ref{tab:multimode_state}. Purple solid contours represent the $2\sigma$ boundaries of the experimental data, while black dashed contours correspond to those of the vacuum state.
    }
\end{figure*}

\begingroup
\renewcommand{\arraystretch}{1.2}
\begin{table*}[htbp]
	\caption{\textbf{Nullifiers and entanglement tests.} For the multipartite entangled states shown in Fig.~\ref{fig:multimode_state}, the associated nullifiers and their measured variances are presented in the middle column. The right column displays the van Loock--Furusawa criteria for multipartite entanglement tests~\cite{van_Loock2003}, confirming the entanglement of the experimentally generated states.}
	\centering
	\begin{ruledtabular}
		\begin{tabular}{@{\hspace{0.5em}}c@{\hspace{0.5em}}|lc@{\hspace{1em}}|@{\hspace{0.5em}}cc@{\hspace{0.5em}}}
			Quantum state & \multicolumn{1}{c}{Nullifier} & Nullifier variance & Entanglement test & Test result \\
			\hline
			Three-mode GHZ state &
			$\begin{array}{ll}
				\hd_{1} =& (\hx_{1} - \hx_{2}) / \sqrt{2} \\
				\hd_{2} =& (\hx_{2} - \hx_{3}) / \sqrt{2} \\
				\hd_{3} =& (\hp_{1} + \hp_{2} + \hp_{3}) / \sqrt{3}
			\end{array}$ & 
            $\begin{array}{c} 0.636 \pm 0.007 \\ 0.642 \pm 0.007 \\ 0.645 \pm 0.007 \end{array}$ &
			$\begin{array}{c@{\hspace{0.25em}}c@{\hspace{0.25em}}}
				\D \hd_{1} + \D \hd_{3} <& 4 / \sqrt{6} \\
				\D \hd_{2} + \D \hd_{3} <& 4 / \sqrt{6}
			\end{array}$ &
			$\begin{array}{c} 1.281 \pm 0.010 \\ 1.286 \pm 0.010 \end{array}$
			\\ \hline
			Four-mode GHZ state &
			$\begin{array}{ll}
				\hd_{1} =& (\hx_{1} - \hx_{2}) / \sqrt{2} \\
				\hd_{2} =& (\hx_{2} - \hx_{3}) / \sqrt{2} \\
				\hd_{3} =& (\hx_{3} - \hx_{4}) / \sqrt{2} \\
				\hd_{4} =& (\hp_{1} + \hp_{2} + \hp_{3} + \hp_{4}) / 2
			\end{array}$ &
            $\begin{array}{c} 0.628 \pm 0.007\\ 0.636 \pm 0.007\\ 0.642 \pm 0.007 \\ 0.645 \pm 0.007 \end{array}$ &
			$\begin{array}{c@{\hspace{0.75em}}c@{\hspace{0.75em}}}
				\D \hd_{1} + \D \hd_{4} <& \sqrt{2} \\
				\D \hd_{2} + \D \hd_{4} <& \sqrt{2} \\
				\D \hd_{3} + \D \hd_{4} <& \sqrt{2}
			\end{array}$ &
			$\begin{array}{c} 1.273 \pm 0.010\\ 1.281 \pm 0.010\\ 1.286 \pm 0.010 \end{array}$ 
			\\ \hline
			$\begin{array}{c} \textrm{Square cluster state} \\ \textrm{(locally equivalent form)} \end{array}$ &
			$\begin{array}{ll}
				\hd_{1} =& (\hx_{1} + \hx_{2} - \hx_{3}) / \sqrt{3} \\
				\hd_{2} =& (\hx_{3} - \hx_{4}) /\sqrt{2} \\
				\hd_{3} =& (\hp_{1} - \hp_{2}) /\sqrt{2} \\
				\hd_{4} =& (\hp_{2} + \hp_{3} + \hp_{4}) / \sqrt{3}
			\end{array}$ &
            $\begin{array}{c} 0.629 \pm 0.007\\ 0.635 \pm 0.007\\ 0.645 \pm 0.007 \\ 0.652 \pm 0.007 \end{array}$ &
			$\begin{array}{c@{\hspace{0.25em}}c@{\hspace{0.25em}}}
				\D \hd_{1} + \D \hd_{3} <& 4 / \sqrt{6} \\
				\D \hd_{2} + \D \hd_{4} <& 4 / \sqrt{6} \\
				\D \hd_{1} + \D \hd_{4} <& 4 / 3
			\end{array}$ &
			$\begin{array}{c} 1.274 \pm 0.010 \\ 1.287 \pm 0.010 \\ 1.280 \pm 0.010 \end{array}$
		\end{tabular}\label{tab:multimode_state}
	\end{ruledtabular}
\end{table*}
\endgroup

\begin{figure*}[htbp]
    \vspace{20mm}
	\includegraphics[width=0.5\linewidth]{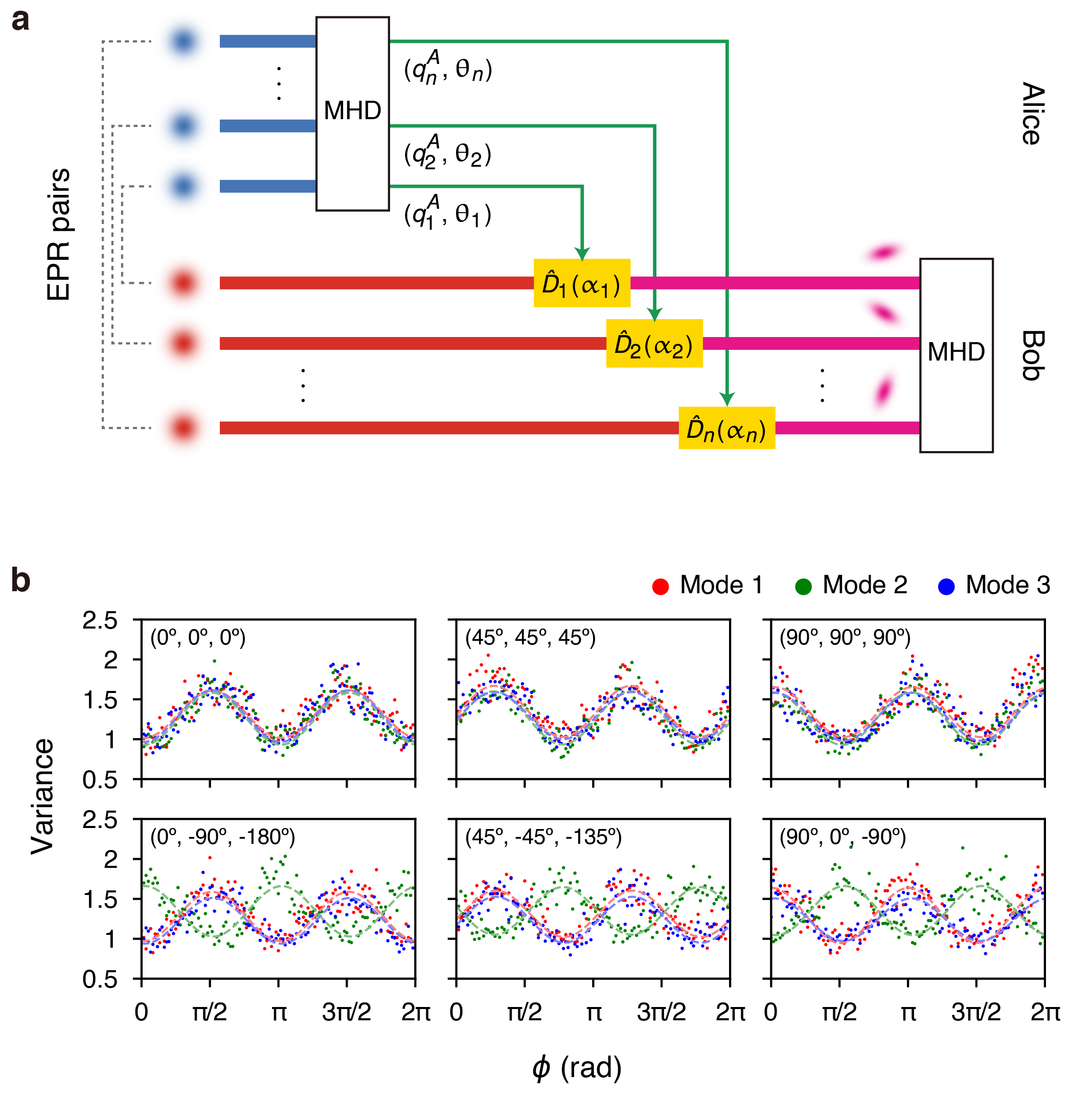}
	\centering
	\caption{\label{fig:cond} \textbf{Conditional preparation of multimode states.} \textbf{a}, Multiple EPR pairs of modes are shared between Alice and Bob. Using a multimode homodyne detector (MHD), Alice obtains multimode quadrature outcomes $({q}_{1}^A,{q}_{2}^A,\ldots,{q}_{n}^A)$ at measurement angles of $(\theta_{1},\theta_{2},\ldots,\theta_{n})$.
    After Bob applies a displacement operator $\hat{D}_k(\alpha_{k})$ at each mode $k$, a multimode squeezed state is prepared and subsequently measured with an MHD for verification. Note that the effect of the displacement operators can also be reproduced by post-processing the quadrature outcomes at Bob.
    \textbf{b}, Quadrature measurement of conditionally prepared three-mode states. Variances of Bob's three modes are plotted as the measurement angles $\phi$ for all modes at Bob are varied simultaneously. A total of 57,600 quadrature outcomes are obtained, and each data point represents the variance calculated from 480 quadrature outcomes obtained within the angle range of $\pm 3^{\circ}$ around each $\phi$. The dashed lines are the sinusoidal fits to the data.
    Alice's measurement angles $(\theta_{1}, \theta_{2}, \theta_{3})$ chosen for conditional state preparation are indicated in the upper-left corner of each plot.}
\end{figure*}

A crucial advantage of multimode homodyne detection is that it provides direct access to quadrature measurements in a superposed mode basis without requiring a beam splitter network. More precisely, one can simultaneously measure either the $\hx$- or $\hp$-quadratures in a superposed basis defined by a real unitary transformation~\cite{Fabre:2020aa, Weedbrook:2012fe}. For example, the superposition $(\hx_1 + \hx_2)/\sqrt{2}$ can be measured by adding the individual outcomes $x_1$ and $x_2$ as $(x_1 + x_2)/\sqrt{2}$. Leveraging this property, we access the multimode quantum light in Fig.~\ref{fig:epr}a in a superposed mode basis \cite{Roh2025, Jia:2025aa} and verify its multipartite entanglement. We investigate three multipartite entangled states: three-mode and four-mode Greenberger--Horne--Zeilinger (GHZ) states~\cite{Su2007} and a square cluster state in a locally equivalent form~\cite{Yukawa2008}.

Figure~\ref{fig:multimode_state}a--c show the mode bases for the three entangled states. We observe strong correlations among quadrature outcomes in different modes, as shown in Fig.~\ref{fig:multimode_state}d--f. In the 3D plots, black dots represent 16,000 measurement results for three different quadratures, while red (blue) dots are 2D projections showing pairwise correlations between two $\hx$ ($\hp$) quadratures. These measurement outcomes from different modes enable the direct verification of entanglement in the generated states. We employ the van Loock--Furusawa criteria for multipartite entanglement tests, which require variance measurements of the nullifiers associated with each target state~\cite{van_Loock2003}. Table~\ref{tab:multimode_state} lists the nullifiers and the corresponding entanglement criteria for the three entangled states (see Supplementary Section \Rom{6} for details). The quadrature outcomes associated with these nullifiers exhibit strong correlations, as shown in Fig.~\ref{fig:multimode_state}g--i. As a result, all nullifier variances fall below the vacuum noise level, and all generated states satisfy the entanglement criteria (see Table~\ref{tab:multimode_state}).

\section{Conditional state preparation}

Lastly, we discuss the application of multimode homodyne detection to conditional preparation of multimode states. In this protocol, a measurement on one part of an entangled state conditionally prepares the remaining part into the desired quantum state~\cite{Laurat:2003jb}. Figure~\ref{fig:cond}a presents a schematic of conditional state preparation, extended to the multimode regime. Alice and Bob share $n$ pairs of Einstein–Podolsky–Rosen (EPR) entangled modes, where Alice measures her modes by quadrature operators $\hq_{k}^{A}(\theta_{k}) = \hx_{k}^{A} \cos\theta_{k} + \hp_{k}^{A} \sin\theta_{k}$ ($k$: pair index, $\theta_{k}$: measurement angle). The measurement outcomes $(q_{1}^{A}, q_{2}^{A}, \ldots q_{n}^{A})$ and the angles $(\theta_{1}, \theta_{2}, \ldots, \theta_{n})$ are sent to Bob, where he applies a displacement operator $\hat{D}_1(\alpha_1) \hat{D}_2(\alpha_2) \cdots \hat{D}_n(\alpha_n)$ in his modes with each amplitude given by $\alpha_{k} = -\gamma_{k} q_{k}^A \exp (-i\theta_{k})$. The coefficient $\gamma_{k}$ is $(V^{a}_k - V^{s}_k) / (V^{a}_k + V^{s}_k)$, determined by the initial squeezing variance $V^{s}_k$ and antisqueezing variance $V^{a}_k$ of each pair $k$. The resulting state in Bob is a multimode squeezed state, where, for each $k$, the squeezing and antisqueezing variances are given by $V^{s}_k{}^{\prime} = 2V^{a}_k V^{s}_k / (V^{a}_k + V^{s}_k)$ and $V^{a}_k{}^{\prime} = \left( V^{a}_k + V^{s}_k \right) / 2$, respectively, with the squeezing angle of $- \theta_{k}$. Detailed derivation of the result is in Supplementary Section \Rom{7}.

We conduct a proof-of-principle demonstration of this protocol using the setup in Fig.~\ref{fig:epr}a. In the experiment, three EPR pairs are used, where Alice and Bob have combined bins in Fig.~\ref{fig:setup}b as their modes, (26--29, 21--24, 16--19) and (32--35, 37--40, 42--45), respectively. At Alice, the measurement angles $(\theta_{1},\theta_{2},\theta_{3})$ are set by using the pulse shaper, and the corresponding quadrature outcomes $(q_{1}^A, q_{2}^A, q_{3}^A)$ are obtained from the associated camera pixels. At Bob, quadrature outcomes $(q_{1}^B, q_{2}^B, q_{3}^B)$ are similarly obtained from the same camera while varying the measurement angles identically $\phi_1=\phi_2=\phi_3=\phi$. The effect of the displacement operators is reproduced by subtracting $\gamma_{k} q_{k}^A \cos (\theta_{k} + \phi_{k})$ from each outcome $q_k^B$. The resulting quadrature variances at Bob are presented in Fig~\ref{fig:cond}b, showing the change of the squeezing angles depending on the measurement angles chosen by Alice.

\section{Conclusion}

Overcoming the scalability challenge in quantum measurement, we demonstrate scalable and efficient homodyne detection enabled by a CCD camera. By engineering a camera-based system for quantum measurement, we achieve simultaneous quadrature measurements of 60 optical modes in the shot-noise-limited regime with high efficiency. These capabilities enable the direct observation of quantum features in multimode light, including multimode squeezing and multiple pairs of entangled modes. Moreover, multimode homodyne detection offers two crucial advantages: mode reconfigurability for detecting multipartite entanglement and conditional state preparation for engineering a multimode state. Importantly, our approach operates directly with free-space quantum light, ensuring robustness against a light propagation loss and a coupling loss. This scalable quantum measurement paradigm paves the way for large-scale quantum information processing~\cite{Asavanant2019, Larsen2019, Aghaee_Rad2025, Roh2025, Zhong:2020aa, Madsen2022, Guo:2019el, Takeda:2019aa, Aoki2009, Armstrong2009, Larsen2021}.

While the present work focuses on a one-dimensional camera array for measuring spectral modes, the approach can be readily extended to two-dimensional arrays for spatial-mode measurements. Such an extension would enable million-scale quantum measurements, with only a few milliwatts of LO power enough to drive the large number of detectors. Complementary metal-oxide-semiconductor (CMOS) and application-specific integrated circuit (ASIC) technologies can further enhance the camera detection rate~\cite{Zia:2023aa}. We envision that this scalable measurement technique will have far-reaching impact across various quantum technologies, including fault-tolerant quantum computing~\cite{Konno:2024aa, Larsen:2025aa, Roh2025}, quantum teleportation~\cite{Sokolov:2001il}, quantum imaging~\cite{Casacio2021, Datta2020}, quantum spectroscopy~\cite{Cai2021, Herman2025, Adamou2025}, and quantum tomography~\cite{Roh2025, Gwak2024}.

\section*{Acknowledgment}
This work was supported by the Ministry of Science and ICT (MSIT) of Korea (RS-2025-00562372, RS-2024-00408271, RS-2024-00442762, RS-2023-NR119925) under the Information Technology Research Center (ITRC) support program (IITP-2026-RS-2020-II201606) and Institute of Information \& Communications Technology Planning \& Evaluation (IITP) grant (RS-2025-25464959, RS-2022-II221029).

\newpage

\providecommand{\noopsort}[1]{}\providecommand{\singleletter}[1]{#1}%

\onecolumngrid
\newpage
\setcounter{figure}{0}
\renewcommand{\thefigure}{E\arabic{figure}}
\renewcommand{\theHfigure}{E\arabic{figure}}

\begin{figure*}[p]
	\includegraphics[width=\linewidth]{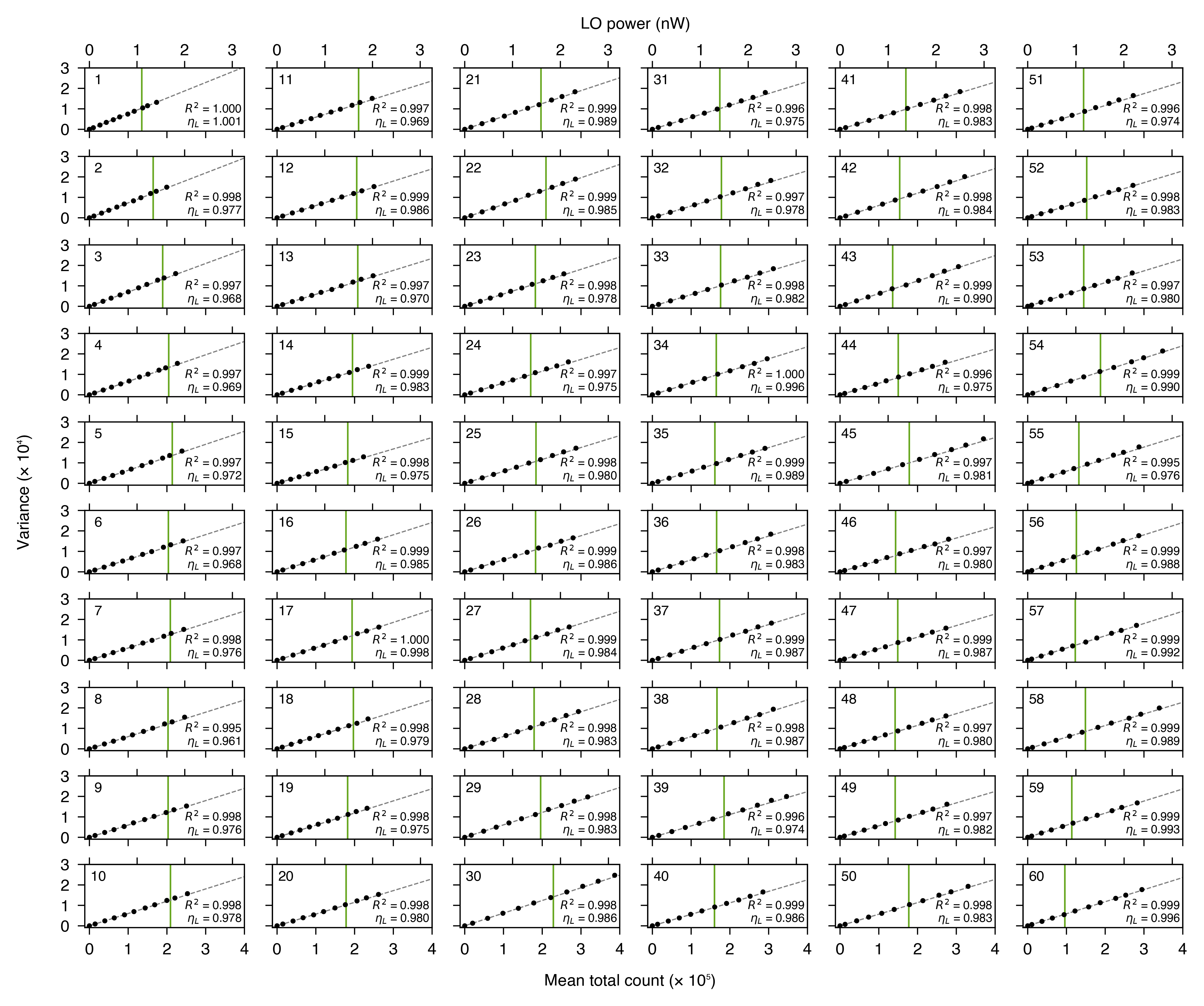}
	\caption{\label{fig:linear_raw} \textbf{Linearity test for 60 frequency-bin modes.} The same method as in Fig.~\ref{fig:setup}d is applied to all 60 modes. Dashed lines show the linear components of the quadratic fits to the data, and green vertical lines mark the LO powers used for multimode homodyne detection. The inset in each plot shows the $R^2$ value (top) and the linear-term contribution $\eta_L$ at the corresponding LO power (bottom).}
\end{figure*}

\begin{figure*}[p]
	\includegraphics[width=\linewidth]{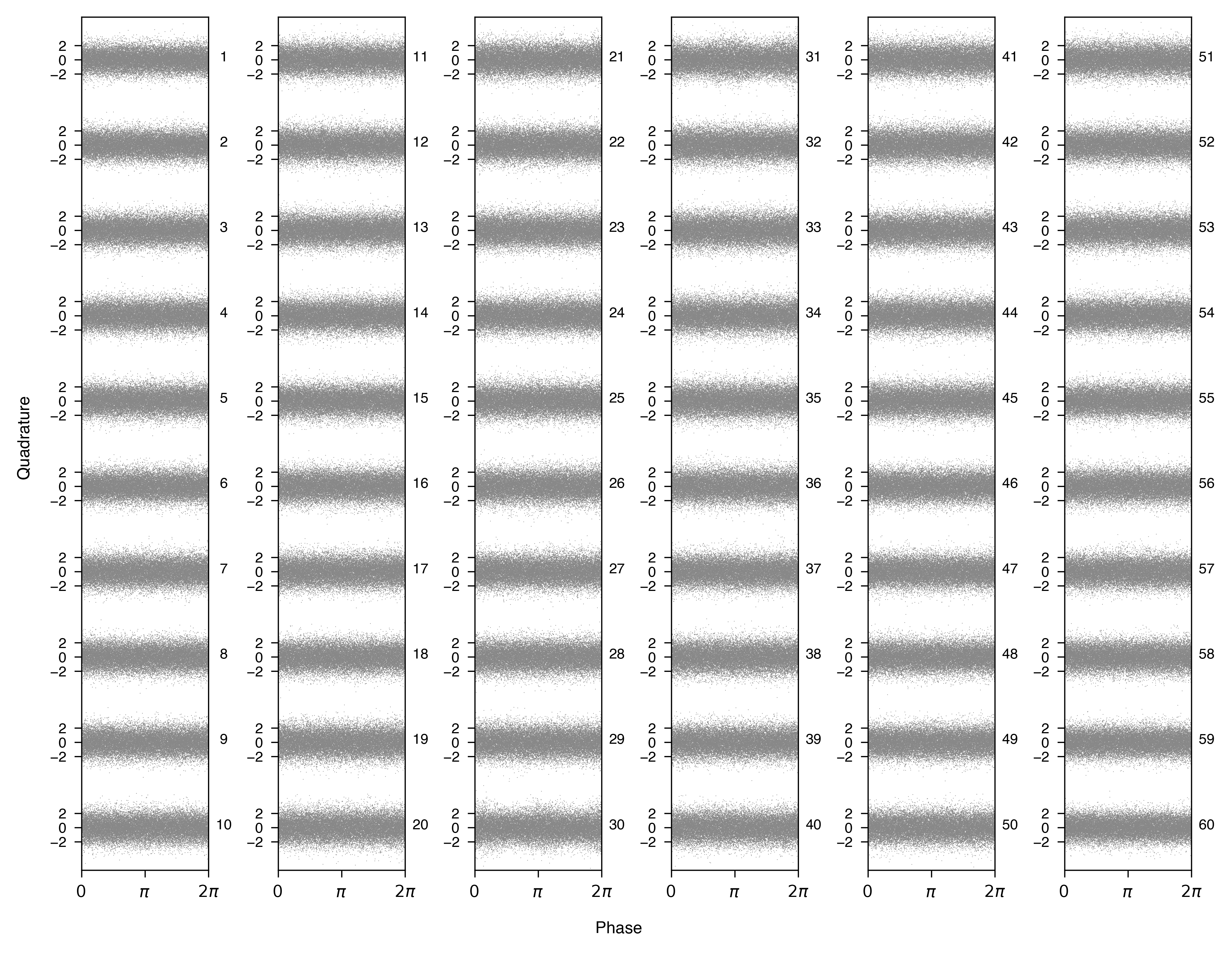}
	\caption{\label{fig:quad_raw} \textbf{Quadrature outcomes for 60 frequency-bin modes.} Quadrature outcomes obtained by the 60-mode homodyne detection in the experiment of Fig.~\ref{fig:epr}a. The corresponding mode index is shown to the right of each plot.}
\end{figure*}
\end{document}